\documentclass[aps,prl,superscriptaddress,twocolumn]{revtex4}
\usepackage{amsmath,bm}
\usepackage{graphicx,epsfig,braket,amssymb,amsfonts,eufrak,tabularx,multirow,amsmath,graphicx,color}
\usepackage{newtxtext}

\newcommand{\bd}{\begin{displaymath}}
\newcommand{\ed}{\end{displaymath}}
\renewcommand{\v}[1]{{\bf #1}}
\newcommand{\bpm}{\begin{pmatrix}}
\newcommand{\epm}{\end{pmatrix}}

\begin{document}

\title{Topological Magnon-Phonon Hybrid Excitations in Two-Dimensional Ferromagnets with Tunable Chern Numbers}

\author{Gyungchoon Go}
%\email{gyungchoon@gmail.com}
\affiliation{Department of Materials Science and Engineering, Korea University, Seoul 02841, Korea}
\author{Se Kwon Kim}
\email{kimsek@missouri.edu}
\affiliation{Department of Physics and Astronomy, University of Missouri, Columbia, Missouri 65211, USA}
\author{Kyung-Jin Lee}
\email{kj_lee@korea.ac.kr}
\affiliation{Department of Materials Science and Engineering, Korea University, Seoul 02841, Korea}
\affiliation{KU-KIST Graduate School of Converging Science and Technology, Korea University, Seoul 02841, Korea}

\begin{abstract}
We theoretically investigate magnon-phonon hybrid excitations in two-dimensional ferromagnets. The bulk bands of hybrid excitations, which are referred to as magnon-polarons, are analytically shown to be topologically nontrivial, possessing finite Chern numbers. We also show that the Chern numbers of magnon-polaron bands and the number of band-crossing lines can be manipulated by an external magnetic field. For experiments, we propose to use the thermal Hall conductivity as a probe of the finite Berry curvatures of magnon-polarons. Our results show that a simple ferromagnet on a square lattice supports topologically nontrivial magnon-polarons, generalizing topological excitations in conventional magnetic systems.
\end{abstract}

\maketitle

\paragraph{Introduction---\hspace{-5pt}}

Since Haldane's prediction of the quantized Hall effect without Landau levels~\cite{Haldane1988},
intrinsic topological properties of electronic bands have emerged as a central theme in condensed matter physics. The band topology can be characterized by emergent vector potential and associated magnetic field defined in momentum space for electron wavefunctions, called Berry phase and Berry curvature, respectively~\cite{Xiao2010}. The Berry curvature is responsible for various phenomena on electron transport such as anomalous Hall effect~\cite{Onoda2006, Sinitsyn2007} and spin Hall effect~\cite{Murakami2003, Kane2005, Bernevig2006}.
%and orbital magnetism~\cite{Thonhauser2005,Xiao2005}.
In addition, nontrivial topology of bulk bands gives rise to chiral or helical edge states according to the bulk-boundary correspondence~\cite{Hatsugai1988}.

Recently, research on the effects of Berry curvature on transport properties, which was initiated for electron systems originally, has expanded to transport of collective excitations in various systems. In particular, magnetic insulators, which gather great attention in spintronics due to their utility for Joule-heat-free devices~\cite{Kajiwara2010}, have been investigated for nontrivial Berry phase effects on their collective excitations~\cite{Katsura2010, Onose2010, Matsumoto2011, Matsumoto2011b, Shindou2013,Zhang2010}: spin waves (magnons) and lattice vibrations (phonons). Previous studies exclusively considering either only magnons or only phonons showed that they can have the topological bands of their own, thereby exhibiting either the magnon Hall effect in chiral magnetic systems~\cite{Katsura2010, Onose2010, Matsumoto2011, Matsumoto2011b, Shindou2013} or the phonon Hall effect~\cite{Zhang2010} when the Raman spin-phonon coupling is present.

Interestingly, the hybridized excitation of magnons and phonons, called a magnetoelastic wave~\cite{Ogawa2015} or magnon-polaron~\cite{KikkawaPRL2016}, is able to exhibit the Berry curvature and thus nontrivial topology due to magnon-phonon interaction~\cite{Takahashi2016, Zhang2019, Park2019}, even though each of magnon system and phonon system has a trivial topology. In noncollinear antiferromagnets, the strain-induced change (called striction) of the exchange interaction is able to generate the nontrivial topology in the magnon-phonon hybrid system~\cite{Park2019}.
In ferromagnets, which are of main focus in this work, nontrivial topology of magnon-polarons is obtained by accounting for long-range dipolar interaction~\cite{Takahashi2016}. In addition, in ferromagnets with broken mirror symmetry, the striction of Dzyaloshinskii-Moriya (DM) interaction leads to topological magnon-polaron bands~\cite{Zhang2019}.
\begin{figure}[t]
\includegraphics[width=86mm]{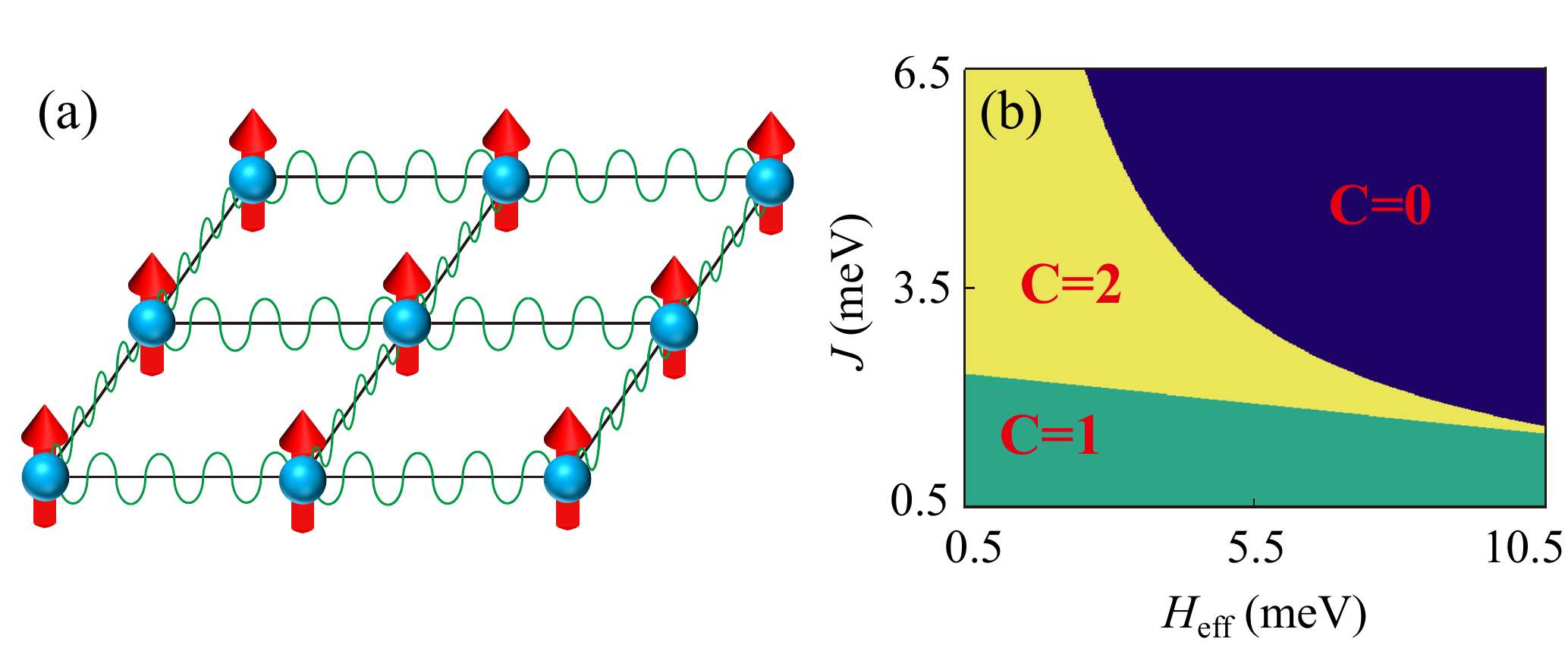}
\caption{(a) The schmematic illustration of the magnon and phonon system. The ground state of the magnetization is given by the uniform spin state along the $z$ axis (red arrow). (b) The Chern number of our magnon-phonon hybrid system.
$H_{\rm eff}$ represents the effective magnetic field including the anisotropy field and the external magnetic field, $H_{\rm eff} = K_z S + \cal B$.
Here we use the parameters $S = 3/2$, $\hbar \omega_0 = 10$ meV, and $M c^2 = 5\times 10^{10}$ eV.
} \label{fig:1}
\end{figure}

In this Letter, we theoretically investigate the topological aspects of the magnon-phonon hybrid excitation in a simple two-dimensional (2D) square-lattice ferromagnet with perpendicular magnetic anisotropy [see Fig.~\ref{fig:1}(a) for the illustration of the system].
Several distinguishing features of our model are as follows. Our model is optimized for atomically thin magnetic crystals, \textit{i.e.}, 2D magnets. The recent discovery of magnetism in 2D van der Waals materials opens huge opportunities for investigating unexplored rich physics and future spintronic devices in reduced dimensions~\cite{McGuire2015, Zhang2015, Lee2016, Gong2017, Huang2017, Bonilla2018, OHara2018, Fei2018, Deng2018, Burch2018, Gibertini2019}.
Because we consider 2D model, we ignore the non-local dipolar interaction, which is not a precondition for a finite Berry curvature in 2D magnets.
Moreover, the Berry curvature we find does not require a special spin asymmetry such as the DM interaction nor a special lattice symmetry: Our 2D model description is applicable for general thin film ferromagnets.
Therefore, we show in this work that even without such long-range dipolar interaction, DM interaction, or special lattice symmetry, the nontrivial topology of magnon-phonon hybrid can emerge by taking account of the well-known magnetoelastic interaction driven by Kittel~\cite{Kittel1958}. As the Kittel's magnetoelastic interaction originates from the magnetic anisotropy, which is ubiquitous in ferromagnetic thin film structures~\cite{Dieny}, our result does not rely on specific preconditions but quite generic. Furthermore, we show that the topological structures of the magnon-polaron bands can be manipulated by effective magnetic fields via topological phase transition.
We uncover the origin of the nontrivial topological bands by mapping our model to the well-known two-band model for topological insulators~\cite{Bernevig2006}, where the Chern numbers are read by counting the number of topological textures called skyrmions of a certain vector in momentum space.
At the end of this Letter, we propose the thermal Hall conductivity as an experimental probe for our theory.

\paragraph{Model---\hspace{-5pt}}

Our model system is a 2D ferromagnet on a square lattice described by the Hamiltonian
\begin{equation}\label{eq:htot}
H = H_{\rm mag} + H_{\rm ph} + H_{\rm mp} \, ,
\end{equation}
where the magnetic Hamiltonian is given by
\begin{align}
H_{\rm mag} = -J \sum_{\langle i,j\rangle} \v S_i\cdot \v S_j - \frac{K_z}{2} \sum_i S_{i,z}^2 - {\cal B} \sum_i S_{i,z},
\end{align}
where $J > 0$ is the ferromagnetic Heisenberg exchange interaction, $K_z >0 $ is the perpendicular easy-axis anisotropy, and $\cal B$ is the external magnetic field applied along the easy axis. Throughout the paper, we focus on the cases where a ground state is the uniform spin state along the $z$ axis: $\mathbf{S}_i = \hat{\mathbf{z}}$. The phonon system accounting for the elastic degree of freedom of the lattice is described by the following Hamiltonian:
\begin{align}\label{Hph}
H_{\rm ph} = \sum_i \frac{\v p_i^2}{2M} + \frac{1}{2} \sum_{i,j,\alpha,\beta} u_i^\alpha \Phi_{i,j}^{\alpha,\beta} u_j^\beta,
\end{align}
where ${\v u}_i$ is the displacement vector of the $i$th ion from its equilibrium position,
$\v p_i$ is the conjugate momentum vector, $M$ is the ion mass, and $\Phi_{i,j}^{\alpha,\beta}$ is a force constant matrix.
The magnetoelastic coupling is modeled by the following Hamiltonian term~\cite{Kittel1958,Thingstad2019}:
\begin{align}\label{mpint}
H_{\rm mp} = \kappa \sum_i \sum_{\v e_i} \left(\v S_i \cdot \v e_i \right) \left(u^z_i - u^z_{i+\v e_i} \right),
\end{align}
where $\kappa$ is the strength of the magnon-phonon interaction and $\v e_i$'s are the nearest neighbor vectors.
Equation~\eqref{mpint} describes the magnetoelastic coupling as a leading order in the magnon amplitude, where the in-plane components of the displacement vector do not appear.

We note here that our model Hamiltonian does not include the dipolar interaction and the DM interaction, distinct from the model considered in Refs.~\cite{Takahashi2016} and \cite{Zhang2019}.
Because the above-mentioned interactions are absent in our model, neither ferromagnetic system nor elastic system exhibits the thermal Hall effect when they are not coupled.
In other words, they are invariant under the combined action of time-reversal ($\cal T$) and spin rotation by $180^\circ$ around an in-plane axis ($\cal C$)~\cite{Zhang2019}.
It is the magnetoelastic coupling term $H_{\rm mp}$ that breaks the combined symmetry $\cal T \cal C$ and thus can give rise to the thermal Hall effect as will be shown below.

\paragraph{Magnon-phonon hybrid excitations---\hspace{-5pt}}

We first diagonalize the magnetic Hamiltonian $H_{\rm mag}$ and the phonon Hamiltonian $H_{\rm ph}$ separately, and then obtain the magnon-phonon hybrid excitations, which are called magnon-polarons, by taking account of the coupling term $H_{\rm mp}$.

The magnetic Hamiltonian is solved by performing the Holstein-Primakoff transformation $S^x_i \approx (\sqrt{2S} / 2) (a_i + a^\dag_i)$, $S^y_i \approx (\sqrt{2S} / 2i) (a_i - a^\dag_i)$, $S^{z}_i = S - a^\dag_i a_i$, where $a_i$ and $a_i^\dagger$ are the annihilation and the creation operators of a magnon at site $i$. By taking the Fourier transformation, $a_i = \sum_{\v k} e^{i \v k\cdot \v R_i} a_{\v k} / \sqrt{N}$, where $N$ is the number of sites in the system, we diagonalize the magnetic Hamiltonian in the momentum space:
\begin{align}
H_{\rm mag} = \sum_{\v k} \hbar \omega_m(\v k) a^\dag_{\v k} a_{\v k},
\end{align}
where the magnon dispersion is given by $\omega_m (\v k) = [2JS\left(2-\cos{k_x} - \cos{k_y}\right) + K_zS + {\cal B}]/\hbar$.

For the elastic Hamiltonian $H_{\rm ph}$, it is also convenient to describe in the momentum space:
\begin{align}
H_{\rm ph} = \sum_{\v k} \left[\frac{{p}^z_{-\v k} {p}^z_{\v k}}{2M} + \frac{1}{2} {u}^z_{-\v k} \Phi(\v k) {u}^z_{\v k}\right],
\end{align}
where only nearest-neighbor elastic interactions are maintained as dominant terms and the momentum-dependent spring constant is $\Phi(\v k) = M\omega_0^2 \left(4 - 2\cos{k_x} - 2\cos{k_y}\right)$, where the characteristic vibration frequency $\omega_0$
corresponds to the elastic interaction between two nearest-neighbor ions. To obtain the quantized excitations of the phonon
system, we introduce the phonon annihilation operator $b_{\v k}$ and the creation operator $b_{\v k}^\dagger$ in such a way that
\begin{align}
\label{eq:d-hmag}
&u^z_{\v k} = \sqrt{\frac{\hbar}{M \omega_{p}(\v k)}} \left(\frac{b_{\v k} + b^\dag_{-\v k}}{\sqrt2}\right),\\
&p^z_{\v k} = \sqrt{\hbar M \omega_{p}(\v k)}  \left(\frac{b_{-\v k} - b^\dag_{\v k}}{{\sqrt2} i}\right),
\end{align}
where the phonon dispersion is given by $\omega_{p}(\v k) = \omega_0 \sqrt{4 - 2\cos{k_x} - 2\cos{k_y}}$. This leads to the following diagonalized phonon Hamiltonian:
\begin{align}
\label{eq:d-hph}
H_{\rm ph} = \sum_{\v k} \hbar \omega_{p}(\v k) \left(b^\dag_{\v k} b_{\v k} + \frac12 \right) \, .
\end{align}
In terms of the magnon and phonon operators introduced above, the magnetoelastic coupling term is recast into the following form in the momentum space: $H_{\rm mp} = H_{\rm mp1} + H_{\rm mp2}$, where
\begin{align}
\label{eq:d-hmp}
&H_{\rm mp1} =\tilde\kappa  \sum_{\v k}\left[ a^\dag_{\v k} b_{\v k}\left(-i \sin k_x + \sin k_y\right)\right] + \text{h.c.} \, ,\\
&H_{\rm mp2} =\tilde\kappa  \sum_{\v k}\left[ a^\dag_{-\v k} b^\dag_{\v k}\left(i \sin k_x - \sin k_y\right) \right]+ \text{h.c.} \, ,
\end{align}
with $\tilde \kappa = \kappa \sqrt{\hbar S / (M \omega_{\v k})}$. Note that $H_{\rm mp1}$ conserves the total particle number, whereas $H_{\rm mp2}$ does not. Because of $H_{\rm mp2}$, the total Hamiltonian takes the Bogoliubov-de-Gennes (BdG) form.
\begin{figure}[t]
\includegraphics[width=86mm]{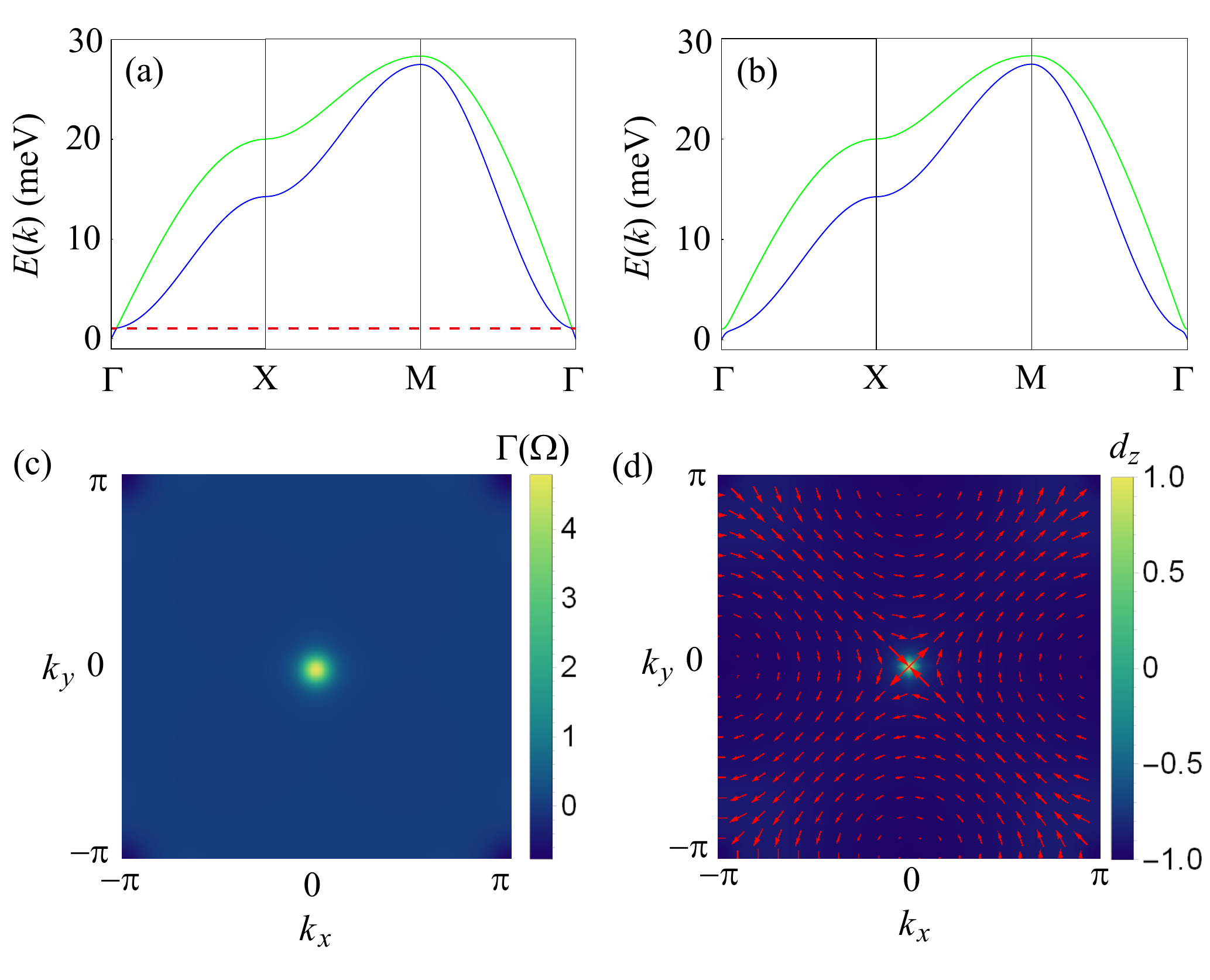}
\caption{The band structure and its topology for $|C| = 1$ case. The band structure for $\kappa = 0$ (a) and $\kappa = 10$ meV/{\AA} (b).
The red dashed line represents the band-crossing points.
(c) Berry curvatures of the upper band in log-scale $\Gamma(\Omega^z) = {\rm sign}(\Omega^z) {\rm log} (1+|\Omega^z|)$ for $\kappa = 10$ meV/{\AA} (d) Schematic illustration of $\v {\hat d} (\v k)$ for $\kappa = 10$ meV/{\AA}.
The in-plane components (${\hat d}_x, {\hat d}_y$) are shown in red arrows.} \label{fig:C1}
\end{figure}

The band structure of magnon-phonon hybrid system is obtained by solving the Heisenberg equations with the above results [Eqs.~\eqref{eq:htot}-\eqref{eq:d-hmp}]
(see the supplementary information for the detailed calculation and the schematic illustration of the band structure~\cite{supple}).
Without magnon-phonon interaction, there are two positive branches consisting of a magnon band and a phonon band.
The two bands cross at $\v k$ points satisfying $\omega_m (\v k)= \omega_p (\v k)$.
Different from the conventional Dirac system, there are innumerable band-crossing points which form a closed line.
These band-crossing lines are removed by the magnon-phonon interaction $\propto \kappa$, which induces the nontrivial topological property of the bands, characterized by the Berry curvatures.
In the BdG Hamiltonian, the Berry curvature is given by~\cite{Zhang2019, Park2019, Cheng2016}
\begin{equation}
\boldsymbol \Omega_n (\v k) = \nabla\times {\v A}_n(\v k) \, ,
\end{equation}
where $\v A_n = i \langle \psi_{n,\v k}|{\cal J}\nabla_{\v k}|\psi_{n,\v k} \rangle$ and $\psi_{n,\v k}$ are the $n$-th eigenstates (see supplementary information for details).
The topological property of the whole system is determined by the Chern number of bands, which is the integral of the Berry curvature over the Brillouin zone~\cite{Qi2008}.
In Fig.~\ref{fig:1}(b), we show the Chern number of our bosonic system with nonzero magnon-phonon interaction $\kappa$.
In our system, the Chern number can be one of three integers (0, 1, and 2) depending on the effective magnetic field $H_{\rm eff} = K_z S + \cal B$ and exchange interaction $J$.
This is one of our central results: The magnon-polaron bands in a 2D simple square-lattice ferromagnet are topologically nontrivial even in the absence of dipolar or DM interaction
and their topological property can be controlled by the effective magnetic field.

\begin{figure}[t]
\includegraphics[width=86mm]{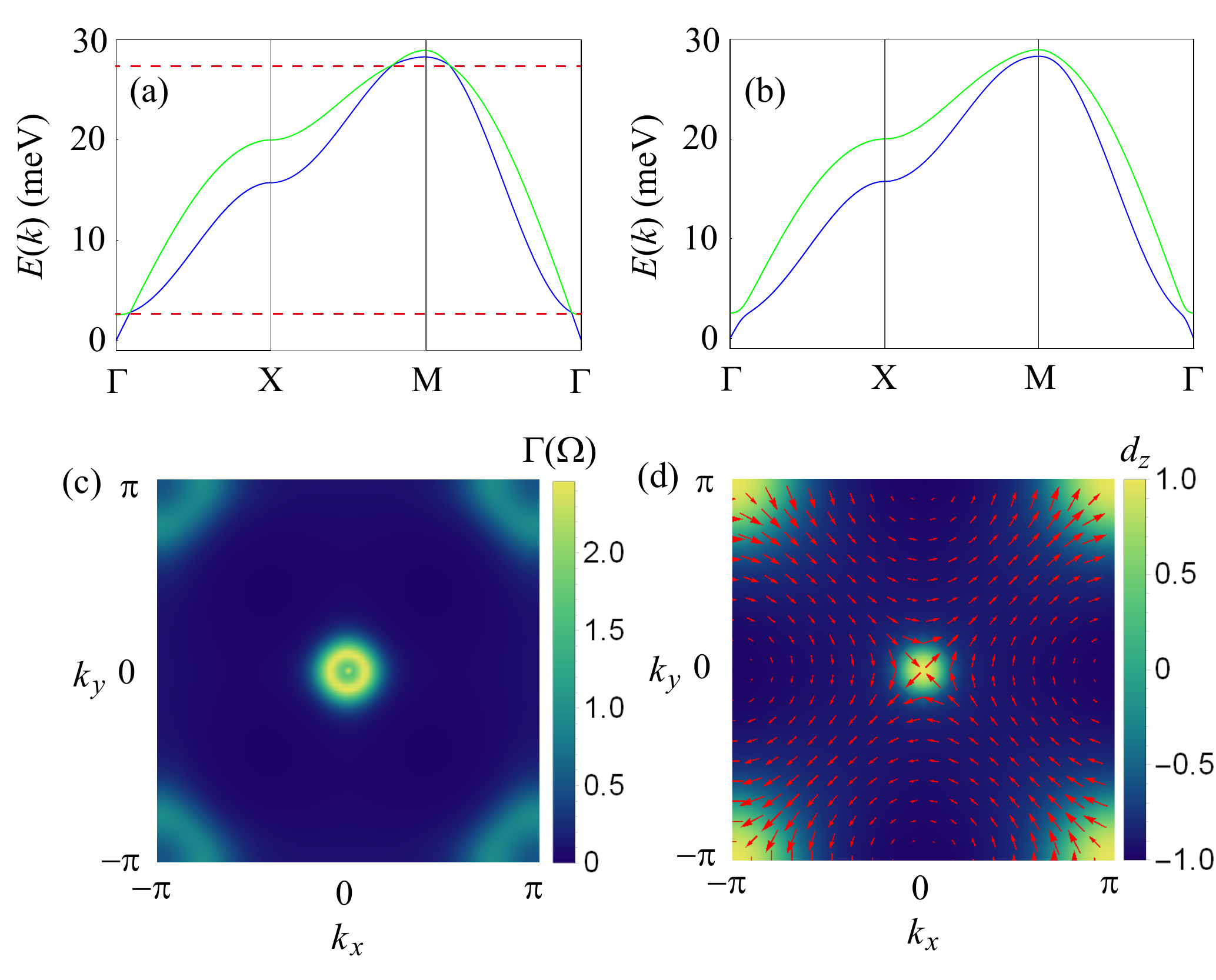}
\caption{The band structure and its topology for $|C| = 2$ case. The band structure for $\kappa = 0$ (a) and $\kappa = 10$ meV/{\AA} (b).
The red dashed line represents the band-crossing points.
(c) Berry curvatures of the upper band in log-scale $\Gamma(\Omega^z) = {\rm sign}(\Omega^z) {\rm log} (1+|\Omega^z|)$ for $\kappa = 10$ meV/{\AA}. (d) Schematic illustration of $\v {\hat d}(\v k)$ for $\kappa = 10$ meV/{\AA}.
The in-plane components (${\hat d}_x, {\hat d}_y$) are shown in red arrows.} \label{fig:C2}
\end{figure}

\paragraph{Origin of the topological property---\hspace{-5pt}}

The origin of the nontrivial magnon-polaron bands obtained above can be understood through the mapping our system to the well-known model for two-dimensional topological insulators such as HgTe~\cite{Bernevig2006, Qi2008}. Considering $H_{\rm mp}$ as a weak perturbation with unperturbed Hamiltonian with well-defined energies of magnons and phonons, the effect of particle-number-nonconserving component $H_{\rm mp2}$ on the band structure is much smaller than that of particle-number-conserving part $H_{\rm mp1}$.
Neglecting $H_{\rm mp2}$, the total Hamiltonian is simplified into a single-particle two-band Hamiltonian
\begin{equation}
H \approx  \sum_{\v k} \left(
                                                                      \begin{array}{cc}
                                                                        a^\dag_\v k & b^\dag_\v k \\
                                                                      \end{array}
                                                                    \right)
 {\cal H_{\v k}}  \left(
                                     \begin{array}{c}
                                       a_\v k \\
                                       b_\v k \\
                                     \end{array}
                                   \right)
 ,
\end{equation}
where
\begin{align}\label{Hsimp}
{\cal H_{\v k}} = \left(
                    \begin{array}{cc}
                      \hbar \omega_{m}(\v k) & \tilde\kappa (\sin k_y - i \sin k_x) \\
                      \tilde\kappa (\sin k_y + i \sin k_x) & \hbar \omega_{p}(\v k) \\
                    \end{array}
                  \right)
.
\end{align}
In terms of the Pauli matrices $\boldsymbol \sigma = (\sigma_x, \sigma_y, \sigma_z)$, we write Eq.~\eqref{Hsimp} in a more compact form
\begin{align}
{\cal H_{\v k}} = \frac\hbar2 \left[\omega_m (\v k) + \omega_p ({\v k})\right] I_{2\times 2} + \v d (\v k)\cdot \boldsymbol \sigma,
\end{align}
where
\begin{align}
\v d (\v k) = \left(\tilde \kappa \sin k_y, \tilde \kappa \sin k_x, \frac\hbar2 \left(\omega_m (\v k) - \omega_p ({\v k})\right)\right).
\end{align}
The band structure for the above Hamiltonian is given by
\begin{align}
E_\pm (\v k) = \frac\hbar2 \left[\omega_m (\v k) + \omega_p ({\v k})\right] \pm |\v d (\v k)|.
\end{align}
In terms of $\v d$ vectors, the Berry curvature is written explicitly as
\begin{equation}
\Omega^z_\pm (\v k) = \mp \frac12 \v {\hat d} (\v k)\cdot\left(\frac{\partial  \v {\hat d} (\v k)}{\partial {k_x}}\times\frac{\partial  \v {\hat d}(\v k)}{\partial {k_y}}\right) \, .
\end{equation}
The corresponding expression for the Chern number is given by~\cite{Qi2008, Volovik1988,Qi2006}
\begin{align}\label{CN}
C_\pm = \frac{1}{2\pi} \int dk_x dk_y \Omega^z_\pm (\v k)\, ,
\end{align}
which is the skyrmion number of the $\v d$ vector~\cite{Qi2008}, counting how many times $\v {\hat d}$ wraps the unit sphere in the Brillouin zone. From Eq.~\eqref{Hsimp}, we read that the magnon band and phonon band cross at $\v k$ points satisfying $\omega_m (\v k)= \omega_p (\v k)$ without the magnon-phonon interaction. These band crossing points are opened by the magnon-phonon interaction $\propto \kappa$ and the finite Berry curvatures are induced near the gap opening region. After integrating the Berry curvatures over the Brillouin zone, we obtain $C_\pm = 0$, $\pm 1$ or $\pm 2$.
The two-band model has almost identical band structures and Berry curvatures to those of full Hamiltonian, where $H_{{\rm mp}2}$ is additionally considered (see the supplemental information).

In Fig.~\ref{fig:C1} and Fig.~\ref{fig:C2}, we show that the bulk band structures and their topological properties for $|C| = 1$ and $|C| = 2$, respectively.
For calculation, we use the parameters of the monolayer ferromagnet ${\rm CrI}_3$ in Ref.~\cite{Huang2017, Zhang2015, Lado2017, Zhang2019} ($J = 2.2$ meV, $K_z = 1.36$ meV, $S = 3/2$,  and $M c^2 = 5\times 10^{10}$ eV). The force constant between the nearest-neighbor phonon is assumed as $\hbar \omega_0 = 10$ meV.
The external magnetic field ${\cal B} = -0.1$ meV is chosen for Fig.~\ref{fig:C1} and ${\cal B} = 0.1$ meV is chosen for Fig.~\ref{fig:C2}.
In Fig.~\ref{fig:C1}(a), we find a band-crossing line (red dashed line) which is removed by the magnon-phonon interaction [Fig.~\ref{fig:C1}(b)].
In this case, a dominant contribution of the Berry curvature comes from vicinity of the $\Gamma$-point [Fig.~\ref{fig:C1}(c)].
An intuitive way to verify the topological nature of the system is the number of skyrmions of the unit vector $\v {\hat d} (\v k)$.
In Fig.~\ref{fig:C1}(d), we find that there is a skyrmion at the $\Gamma$-point corresponding to $|C| = 1$.
By changing the sign of external magnetic field $\cal B$, we can modify the band structure with two band-crossing lines [Fig.~\ref{fig:C2}(a)].
In this case, the dominant contribution of the Berry curvature comes from vicinity of the $\Gamma$- and $\rm M$-points [Fig.~\ref{fig:C2}(c)].
In terms of $\v {\hat d} (\v k)$, we find that one skyrmion is located at $\Gamma$-point and the other skyrmion is at $\rm M$-point corresponding to $|C| = 2$ [Fig.~\ref{fig:C2}(d)].

\paragraph{Thermal Hall effect---\hspace{-5pt}}

The finite Berry curvatures of magnon-phonon hybrid excitations give rise to the intrinsic thermal Hall effect as shown below. The semiclassical equations of motion for the wave packet of magnon-phonon hybrid are given by~\cite{Sundaram1999,Xiao2010}
\begin{align}
\dot {\v r}_n = \frac{1}{\hbar} \frac{\partial E_n(\v k)}{\partial \v k} - \dot {\v k} \times {\boldsymbol \Omega}_n(\v k), \quad \hbar \dot{\v k} = - \nabla U(\v r),
\end{align}
where $U(\v r)$ is the potential acting on the wave packet which can be regarded as a confining potential of the bosonic excitation.
Near the edge of sample, the gradient of the confining potential produces the anomalous velocity, $\nabla U(\v r) \times {\boldsymbol \Omega}_n(\v k)$. In equilibrium, the edge current circulates along the whole edge and net magnon current is zero along any in-plane direction.
However, if the temperature varies spatially, the circulating current does not cancel, which causes the thermal Hall effect~\cite{Matsumoto2011}.

\begin{figure}[t]
\includegraphics[width=86mm]{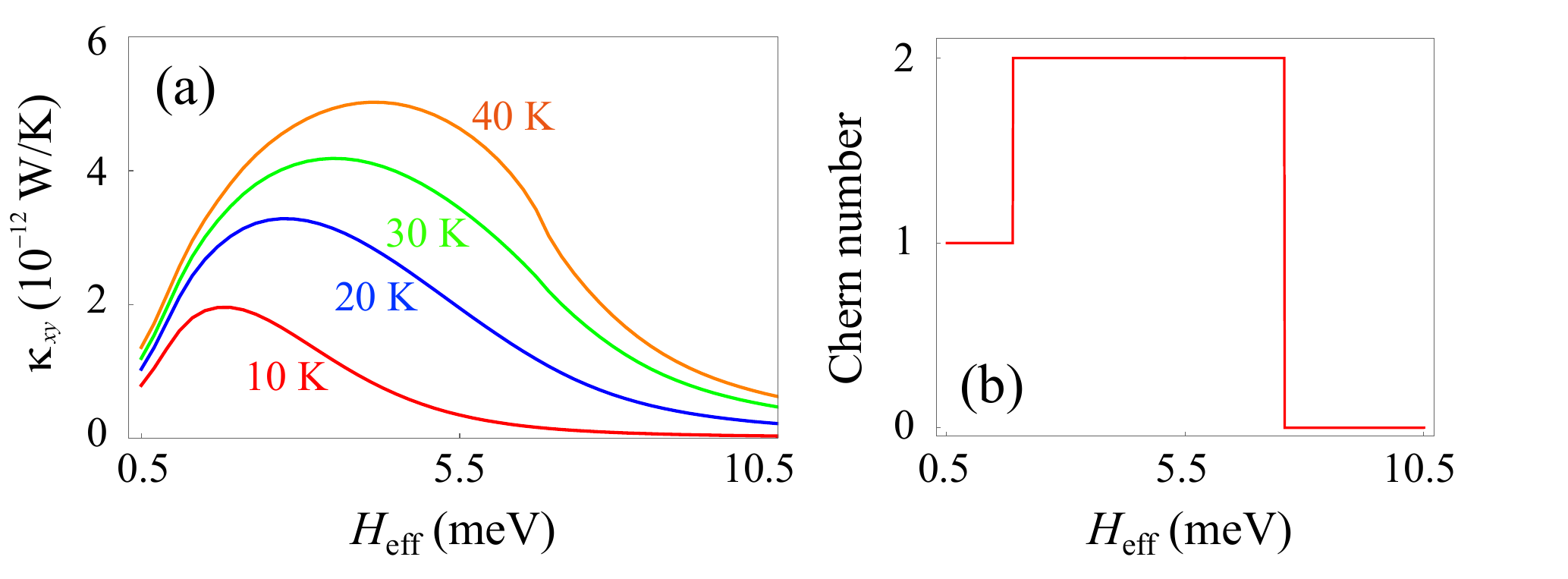}
\caption{(a) Dependence of thermal Hall conductivity on the effective magnetic field $H_{\rm eff}$ and with different temperatures $T$ using the parameters in the main text.
(b) Dependence of Chern number on the effective magnetic field $H_{\rm eff}$  using the parameters in the main text.}\label{fig:THC1}
\end{figure}

The Berry-curvature-induced thermal Hall conductivity is given by~\cite{Matsumoto2011, Matsumoto2011b}
\begin{align}
\kappa^{xy} = -\frac{k_B^2 T}{\hbar V} \sum_{n, \v k} c_2(\rho_{n,\v k}) \Omega_n^z(\v k),
\end{align}
where $c_2(\rho) = (1+\rho) \ln^2 [(1+\rho) / \rho] - \ln^2 \rho - 2 {\rm Li}_2(-\rho)$, $\rho_{n,\v k} = [e^{(E_n(\v k))/{k_B T} - 1}]^{-1}$ is the Bose-Einstein distribution function with a zero chemical potential,
$k_B$ is the Boltzmann constant, $T$ is the temperature, and ${\rm Li}_2 (z)$ is the polylogarithm function.
In Fig.~\ref{fig:THC1}(a), we show the dependence of thermal Hall conductivity on the effective magnetic field $H_{\rm eff}$ at different temperatures $T$.
For small $H_{\rm eff}$, the thermal Hall conductivity increases with increasing $H_{\rm eff}$. However, for large $H_{\rm eff}$, it decreases with increasing $H_{\rm eff}$.
This behavior of the thermal Hall conductivity can be understood through the Chern number of magnon-polaron bands depicted in Fig.~\ref{fig:THC1}(b).
The absolute value of the Chern number is 1 for small $H_{\rm eff}$, then it jumps up to 2 for a certain value of $H_{\rm eff}$ and vanishes for large $H_{\rm eff}$.

\paragraph{Discussions---\hspace{-5pt}}

In this paper, we investigate the topology of the magnon-polaron bands in a simple 2D ferromagnet without long-range dipolar interaction and DM interaction.
In our model, the topological structure can be controlled by the effective magnetic field which changes the number of band-crossing lines.
Using a perturbation approach, we develop a two-band model Hamilitonian which provides an intuitive understanding of the topological structure of the model.
In the two-band model, the nontrivial topology of the magnon-polaron bands are reflected in the skyrmion number of $\v d(\v k)$ in momentum space.
As an experimental demonstration, we propose that the thermal Hall conductivity arises from the non-trivial topology of the magnon-polaron bands.
The thermal Hall conductivity depends on the effective magnetic field which can be manipulated by the external magnetic field or voltage-induced magnetic anisotropy change~\cite{Maruyama2009}.
Our results show that the magnetoelastic interaction generates nontrivial topology in simple 2D ferromagnets with topological tunability, suggesting the ubiquity of topological transports in conventional magnetic systems with reduced dimensions.

\acknowledgments K.-J. L. acknowledges a support by the National Research Foundation (NRF) of
Korea (NRF-2017R1A2B2006119). G.G. acknowledges a support by the NRF of Korea (NRF-2019R1I1A1A01063594).
S.K.K. was supported by the startup fund at the University of Missouri.

\end{document}